# Explanation of displacement current in a vacuum

Author: Petr Šlechta, Větrná 8, České Budějovice, Czech Republic
Date: 2006 September 22

## 1 Abstract

It is considered that the time derivative of the electric intensity in the Maxwell-Ampere law (displacement current) denotes that a change of electric field generates a magnetic field. This paper shows that there is no reason think a change of electric field generates a magnetic field and the displacement current term has different meaning – it is necessary to be aware of distant conductors and their magnetic fields.

## 2 Introduction

It is not too clear why there is the term $\frac{\partial \vec{D}}{\partial t}$ (displacement current) in the Maxwell-Ampere law $\nabla \times \vec{H} = \frac{\partial \vec{D}}{\partial t} + \vec{j}$. There is no problem with the part $\frac{\partial \vec{P}}{\partial t}$ caused by change of polarization. I want to talk about the part $\varepsilon_0 \frac{\partial \vec{E}}{\partial t}$ that is present even in a vacuum. In my school-book is written: „Displacement current in a vacuum is the first of quantities we have no mechanical analogy for." Also Wikipedie thinks displacement current in a vacuum is a physical phenomenon: „The first part is present everywhere, even in a vacuum; it does not involve any actual movement of charge, but still has an associated magnetic field, as if it were an actual current." And all teachers I know say the same.

The intension of this paper is to show that displacement current in a vacuum creating magnetic field is not necessary to explain the term $\varepsilon_0 \frac{\partial \vec{E}}{\partial t}$ in the Maxwell-Ampere law. It is sufficient to consider only the magnetic field created by conductors and the term $\varepsilon_0 \frac{\partial \vec{E}}{\partial t}$ only informs us about currents that are far from the place where the magnetic field is measured.

## 3 Maxwell-Ampere law

Now we will try to derive the Maxwell-Ampere law and show we need the term $\varepsilon_0 \frac{\partial \vec{E}}{\partial t}$ in it even though only the magnetic field from conductors is considered. We will assume that changes of currents and fields are slow and the environment is a vacuum, so $\varepsilon = \varepsilon_0$ and $\mu = \mu_0$.

### 3.1 Infinitely long direct conductor

Let's have an infinitely long direct conductor with a constant current $I$ and an area $A$ with its boundary $\partial A$ (see Fig.1).

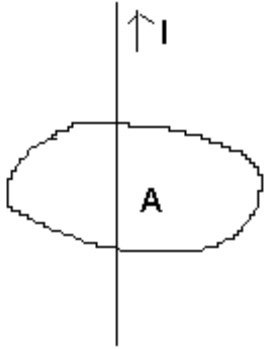

**Fig. 1**

It can be shown by integration that $\oint_{\partial A} \vec{H} d\vec{s} = \iint_A \vec{j} d\vec{A}$. However, this relation cannot be used universally. It works only for an infinitely long direct conductor.

## 3.2 Discontinuous conductor

Now we will try to mystify our equation from the chapter 3.1. We will assume the same configuration but the conductor will be cut off along the area *A*. The cut will be very narrow, so it will form a capacitor (see Fig.2).

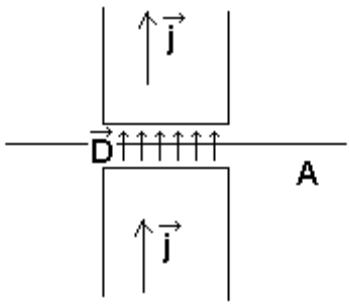

**Fig. 2**

For the electric field in the cut we can write $\dfrac{\partial \vec{D}}{\partial t} = \vec{j}$.

Now we can ask whether our equation $\oint_{\partial A} \vec{H} d\vec{s} = \iint_A \vec{j} d\vec{A}$ is still valid. Value of the left side of the equation has not changed because the magnetic field stays the same (the cut is very narrow). Conversely, the value of the right side of the equation is now 0 since no current flows through the area *A*. However, we have electric field in the cut instead of the current. Therefore, we will add the term $\dfrac{\partial \vec{D}}{\partial t}$ that shows there is some current near that creates magnetic field. The equation will be $\oint_{\partial A} \vec{H} d\vec{s} = \iint_A (\dfrac{\partial \vec{D}}{\partial t} + \vec{j}) d\vec{A}$ in the integral form or

$\nabla \times \vec{H} = \dfrac{\partial \vec{D}}{\partial t} + \vec{j}$ in the differential form.

Now we have the final version of the Maxwell-Ampere law but only for an infinitely long direct conductor (possibly discontinuous). It is necessary to show that the equation is valid for any configuration of conductors. To do this (using the superposition principle) it is sufficient to show that Maxwell-Ampere law is valid around a very short (elementary) conductor because any configuration can be assembled from very short conductors and very long direct conductors. We will show this in the next chapter.

## 3.3 Very short conductor

Since we can use any coordinate system we will assume there is a very short conductor of the length $l$ in the origin oriented along the z-axis. Current $I$ flows through the conductor. Let's assign $\vec{l} = l \cdot \vec{z}_0$ the length and orientation of the conductor. At both the ends of the conductor an electric charge cumulates and it creates an electric dipole moment $\vec{p} = Q \cdot \vec{l}$ with the value $\frac{d\vec{p}}{dt} = I \cdot \vec{l}$. Now we will try to calculate the electric and magnetic field at the position $\vec{r} = [x, y, z]$ (not in the origin) and check if the Maxwell-Ampere law is valid.

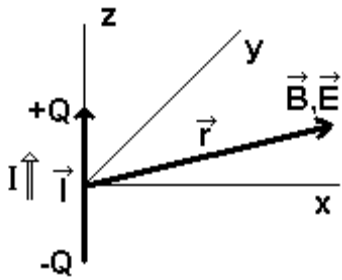

**Fig. 3**

### 3.3.1 Magnetic field

Using Biot-Savart-Laplace law we find the magnetic induction $\vec{B} = \frac{\mu I}{4\pi r^3} \cdot \vec{l} \times \vec{r} = \frac{\mu I l}{4\pi r^3} \cdot [-y, x, 0]$. The magnetic intensity is $\vec{H} = \frac{\vec{B}}{\mu} = \frac{Il}{4\pi r^3} \cdot [-y, x, 0]$. So its curl is $\nabla \times \vec{H} = \frac{Il}{4\pi r^5} \cdot [3xz, 3yz, 3z^2 - r^2]$.

### 3.3.2 Electric field

The electric dipole generates electric field with the potencial $V = \frac{\vec{p} \cdot \vec{r}}{4\pi\varepsilon r^3} = \frac{p}{4\pi\varepsilon} \cdot \frac{z}{r^3}$. Electric intensity is $\vec{E} = -\nabla V = \frac{p}{4\pi\varepsilon r^5} \cdot [3xz, 3yz, 3z^2 - r^2]$. Electric induction is $\vec{D} = \varepsilon \vec{E} = \frac{p}{4\pi r^5} \cdot [3xz, 3yz, 3z^2 - r^2]$. Using the formula $\frac{d\vec{p}}{dt} = I \cdot \vec{l}$ we find that $\frac{\partial \vec{D}}{\partial t} = \nabla \times \vec{H}$.

Since out of the origin there is no electric current ($\vec{j} = 0$) we can write $\nabla \times \vec{H} = \frac{\partial \vec{D}}{\partial t} + \vec{j}$.

# 4 Conclusion

It has been shown that in a vacuum stands $\nabla \times \vec{H} = \frac{\partial \vec{D}}{\partial t} + \vec{j}$ (the Maxwell-Ampere law). We considered only the magnetic field created by conductors accordingly to Biot-Savart-Laplace law. We did not consider that a change of electric field generates magnetic field. So the existence of the term $e_0 \frac{\partial \vec{E}}{\partial t}$ in the Maxwell-Ampere law does not denote any new physical phenomenon (e.g. vacuum polarization) and there is no reason to think that a change of electrical intensity generates magnetic field.